\documentclass[
aps,
prb,
showpacs
amsmath,
amssymb,
preprint,
reprint,
floatfix,
letterpaper,
lengthcheck,
showpacs,
showkeys
]{revtex4-2}

\usepackage{graphicx}
\usepackage{rotating}
\usepackage{amsmath}
\usepackage{bbm}
\usepackage{subfigure}
\usepackage{comment}
\usepackage{color}
\usepackage{braket}
\usepackage{hyperref}
\graphicspath{
  {figures/pdf/}
  {figures/eps/}
}

\begin{document}

\title{Quantum electrodynamics in high harmonic generation: multi-trajectory Ehrenfest and exact quantum analysis}

\author{Sebastian de-la-Pe\~na\textsuperscript{1}}
  \email[Electronic address:\;]{sebastian.delapena@mpsd.mpg.de}

\author{Ofer Neufeld\textsuperscript{1}}

\author{Matan Even Tzur\textsuperscript{2}}

\author{Oren Cohen\textsuperscript{2}}

\author{Heiko Appel\textsuperscript{1}}

\author{Angel Rubio\textsuperscript{1,3}}
    \email[Electronic address:\;]{angel.rubio@mpsd.mpg.de}

\affiliation{\textsuperscript{1} Max Planck Institute for the Structure and Dynamics of Matter, Luruper Ch 149, 22761 Hamburg, Germany}
\affiliation{\textsuperscript{2} Department of Physics and Solid State Institute, Technion–Israel Institute of Technology, Haifa, Israel}
\affiliation{\textsuperscript{3} Center for Computational Quantum Physics, The Flatiron Institute, New York, NY, USA}

\date{\today}

%%%%%%%%%%%%%%%%%%%%%%%%%%%%%%%%%%%%%%%%%%%%%%%%%%%%%%%%%%%%%%%%%%
%                            Abstract                            %
%%%%%%%%%%%%%%%%%%%%%%%%%%%%%%%%%%%%%%%%%%%%%%%%%%%%%%%%%%%%%%%%%%
\begin{abstract}
\noindent High-harmonic generation (HHG) is a nonlinear process in which a material sample is irradiated by intense laser pulses, causing the emission of high harmonics of the incident light. HHG has historically been explained by theories employing a classical electromagnetic field, successfully capturing its spectral and temporal characteristics. However, recent research indicates that quantum-optical effects naturally exist, or can be artificially induced, in HHG, such as entanglement between emitted harmonics. Even though the fundamental equations of motion for quantum electrodynamics (QED) are well-known, a unifying framework for solving them to explore HHG is missing. So far, numerical solutions employed a wide range of basis-sets, methods, and untested approximations. Based on methods originally developed for cavity polaritonics, here we formulate a numerically accurate QED model consisting of a single active electron and a single quantized photon mode. Our framework can in principle be extended to higher electronic dimensions and multiple photon modes to be employed in {\it ab initio} codes for realistic physical systems. We employ it as a model of an atom interacting with a photon mode and predict a characteristic minimum structure in the HHG yield vs. phase-squeezing. We find that this phenomenon, which can be used for novel ultrafast quantum spectroscopies, is partially captured by a multi-trajectory Ehrenfest dynamics approach, with the exact minima position sensitive to the level of theory. On the one hand, this motivates using multi-trajectory approaches as an alternative for costly exact calculations. On the other hand, it suggests an inherent limitation of the multi-trajectory formalism, indicating the presence of entanglement and true quantum effects (especially prominent for atomic and molecular resonances). Our work creates a road-map for a universal formalism of QED-HHG that can be employed for benchmarking approximate theories, predicting novel phenomena for advancing quantum applications, and for the measurements of entanglement and entropy. 
%We show that multi-trajectory Ehrenfest proves to be a useful tool to gain insight from HHG with quantum light and we suggest improvements to these algorithm to cover more sophisticated systems for light, with a multimode implementation, and for matter in a realistic molecular setup.
\end{abstract}

%\pacs{71.15.-m, 31.70.Hq, 31.15.ee}
% 71.15.-m Methods of electronic structure calculations
% 31.70.Hq Time-dependent phenomena: excitation and relaxation
% 31.15.ee Time-dependent density functional theory

\date{\today}

\maketitle

%%%%%%%%%%%%%%%%%%%%%%%%%%%%%%%%%%%%%%%%%%%%%%%%%%%%%%%%%%%%%%%%%%
%                        Introduction                            %
%%%%%%%%%%%%%%%%%%%%%%%%%%%%%%%%%%%%%%%%%%%%%%%%%%%%%%%%%%%%%%%%%%
\section{Introduction}
\label{sec: intro}

%% Context

High-harmonic generation (HHG) is a nonlinear optical process in which molecules~\cite{application-gas-1, application-gas-2}, liquids~\cite{application-liquid}, or solids~\cite{application-solid} are exposed to an intense light source and radiate higher harmonics of the driving light main frequency. This phenomenon has enabled the birth of new research areas like attosecond spectroscopy~\cite{attosecond-physics, attosecond-metrology}, and is routinely used for generating coherent X-rays table-top~\cite{x-ray-1}. Initially, HHG in atomic and molecular systems was understood as a consequence of the semiclassical motion of the electron around the nucleus~\cite{plasma-strong-field} and later it was explained using quantum mechanical models~\cite{Theory-hhg, keldysh-ionization, PRXHHG}. In these models the light source was treated as a classical electromagnetic field, making it inapplicable for analyzing HHG from emerging quantum light sources or for explaining quantum-optical effects and interferometry in the emitted harmonic spectra. \\

% gases liquids and solids
% attosecond spectroscopy
% description of hhg -> semiclassical, quantum mechanical (with classical light)

%% Current results

\noindent Recent research shows that quantum-optical effects are in fact potentially prominent in high-harmonic generation~\cite{quantum-optical-signatures, high-order-harmonics-photon-statistics}. This has motivated the development of new theories~\cite{qo-nature-hhg, schrodinger-cat} capable of accounting for nonclassical light sources~\cite{BSV-husimi, High-harmonic-generation-from-an-atom-in-a-squeezed-vacuum-environment, matan-coherent-squeezed, motion-charged-particles-BSV, generation-squeezed-harmonics} or entanglement in the emitted harmonics, even when the source is treated classically~\cite{Entanglement-squeezing, H2+, strongly-driven-many-body, quantum-optical-description-hhg, squeezing-from-resonance}. The field is also experiencing major experimental efforts~\cite{quantum-optical-signatures, high-order-harmonics-photon-statistics}, e.g. applying quantum sources for HHG~\cite{experiment-BSV, Correlation-function-experiment} or exploring violation of Cauchy-Schwarz's inequalities in the emitted light~\cite{evidence-quantum-hhg}. \\ 

%on - also mention here the original tzallas paper please. it should be first that is mentioned in the exp. part.

%% Context and state-of-the-art

%% Angel comment: I-Te's functional for photon correlation

\noindent The main equations describing such phenomena are known exactly from quantum-electrodynamics (QED)~\cite{cohen-qed, principle-nonlinear-optics-mukamel}. Nonetheless, they cannot be practically solved without resorting to approximations of either the Hamiltonian or wave functions~\cite{quantum-phenomena-attosecond-science}, which has already lead to some developments in cavity materials engineering by making use of density functional theory~\cite{QEDFT-rugge, QEDFT-rugge-2, QEDFT-rugge-3}. However, to date, multiple papers predicted a variety of phenomena based on various methodologies and approximations, mostly ad-hoc, and some not necessarily agreeing with each other. An {\it ab initio} solution of the HHG system is unfeasible even with only a single active electron due to the exponential scaling of the bosonic basis set for highly populated photon states. Such large number of photon modes are essential in HHG that is driven by very intense lasers and causes emission of a broad spectrum. Generally, one would want to exploit the success of semiclassical multi-trajectory techniques in the field of quantum chemistry for the electron-phonon coupling~\cite{MTEF-phonon-1, phonon-multitrajectory}, a formalism which has also been tested for electron-photon systems in the context of spontaneous emission~\cite{spontaenous-emission-multitrajectory, benchmarking}, and employ such an approach for describing quantum HHG. The multi-trajectory Ehrenfest dynamics (MTEF) approach should capture qualitative dynamics intuitively and presents a linear scaling with the system size, bridging the notions of classical electrodynamics with quantum optics and enabling its use for more complicated systems. Yet, it fails to provide an exact quantitative description of processes, e.g. wrong predictions of final state population in spontaneous emission processes~\cite{spontaenous-emission-multitrajectory, benchmarking} or zero-point energy leakage~\cite{zero-energy-leakage}. In the context of HHG, no testing of MTEF of trajectory-based theories has previously been done and that level of approximation is untested.  \\

%% New comment: which paper are you thinking on?

%% Contributions of this paper

\noindent Here we theoretically study HHG in a 1D atom model irradiated by an intense quantum light source. In order to introduce quantum-optical states of light for HHG, we couple the electron to the light field through two models: (i) an exact quantized single photon mode with the frequency of the HHG driving field, leading to a formally accurate quantum model of the dynamics; and (ii) MTEF that approximates the quantized photon mode via multiple semiclassical simulations sampling a quantum-optical distribution function. Both numerical methods can in principle be extended and employed in a universal theory for benchmarking approximations and makings predictions, either by including many electrons into consideration in quantum chemistry codes, or by adding multiple photon modes. We employ these methods in a 1D model atom and test the viability of MTEF by comparing HHG driven by squeezed-coherent light with different degrees of squeezing between both methods. We observe a characteristic minimum emerges in all harmonic orders vs. the phase squeezing parameter, a phenomenon that MTEF only partially captures, potentially exposing true quantum effects in the light-matter entanglement which MTEF systematically neglects. Our work paves the path for the development of an {\it ab initio} framework for solving QED-HHG, and provides a quantitative prediction of the squeezing-dependence of the HHG spectrum that can be used for novel quantum ultrafast spectroscopy and benchmark previous approaches. \\

\begin{figure}[ht]
    \centering
    \includegraphics[width=\linewidth]{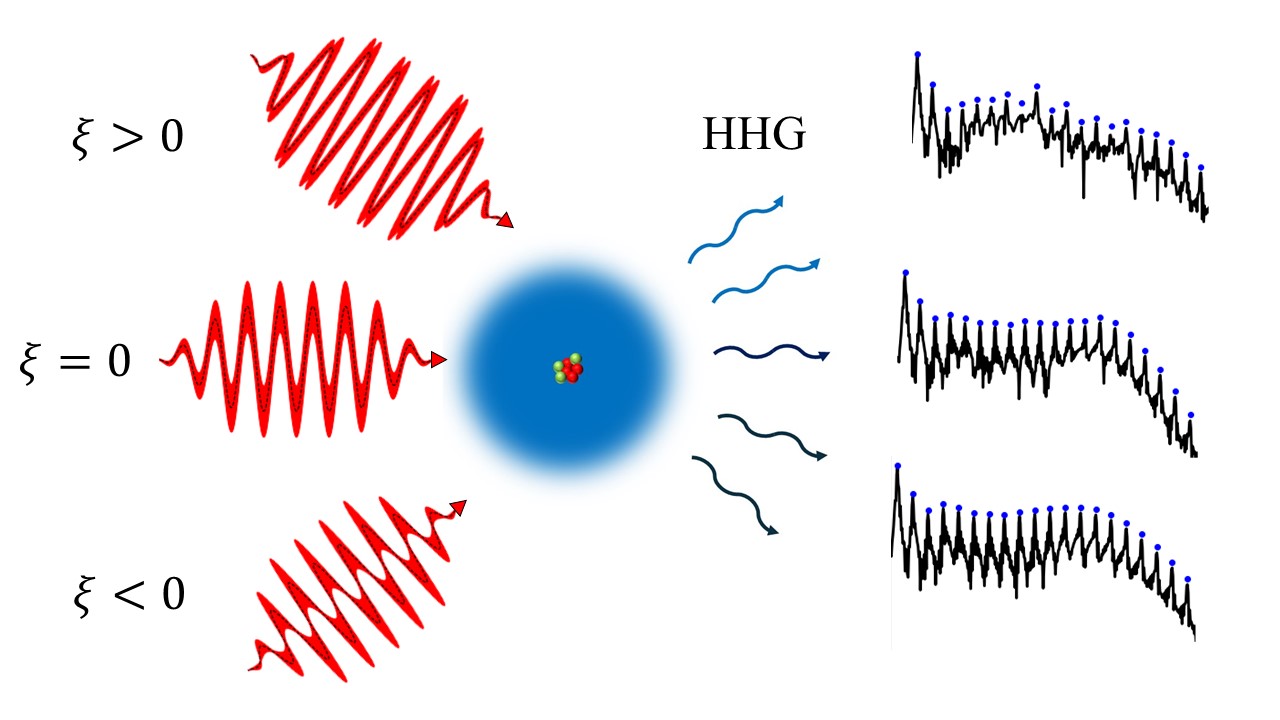}
    \caption{Schematic illustration of HHG driven by quantum light, and leading to quantum-optical effects. Laser fields with different squeezings (left) are irradiated onto an atom and this produces the HHG spectrum with potentially quantum features.}
    \label{fig: figure-paper}
\end{figure}

\noindent The manuscript is ordered as follows: in Sec.~\ref{sec: methods} we describe our theoretical approaches. The comparison between methods for quantum HHG is given in Sec.~\ref{sec: results}, as well as a discussion of the results. Finally, Sec~\ref{sec: conclusion} summarizes our results and presents a future outlook.

\section{Computational methodology}
\label{sec: methods}

Let us begin by describing our main observable of interest, how it is extracted from calculations and our motivation. The standard theory for HHG uses a coherent light pulse with field intensity $\alpha_0$: $\ket{\alpha_0} \equiv \hat{D}(\alpha_0) \ket{0}$, where $\hat{D}$ is the displacement operator~\cite{Scully_Zubairy_1997} and $\ket{0}$ is the vacuum state of the driving photon mode. The main result of the standard HHG theory allows us to compute the electromagnetic emission from the Fourier transform of the electronic dipole $d\mathcal{E}/d\omega \sim |d(\omega)|^2$~\cite{PRXHHG, qo-nature-hhg, jackson-em}. The dipole $d(t)$ is governed by a time-dependent Schr\"odinger equation (TDSE) with a driving electric field $E_{\mathrm{cl}}(t) \equiv \bra{\alpha_0} \hat{E}(t) \ket{\alpha_0} \sim \alpha_0 \cos\left( \omega_L t \right)$, where $\omega_L$ is the laser frequency and the electric field operator $\hat{E}(t)$ is in the interaction picture~\cite{cohen-qed, principle-nonlinear-optics-mukamel}. In this theory the mean field of the driving photon system determines completely the HHG spectrum and, consequently, nonclassical states of light like squeezed states $\ket{\alpha_0, \xi} \equiv \hat{D}(\alpha_0) \hat{S}(\xi) \ket{0}$ (where $\hat{S}$ is the squeezing operator~\cite{Scully_Zubairy_1997} and $\xi$ the degree of squeezing) would lead to the same spectrum owing to their irrelevance in the expectation value of the electric field $E_{\mathrm{cl}}(t) = \bra{\alpha_0} \hat{E}(t) \ket{\alpha_0} = \bra{\alpha_0, \xi} \hat{E}(t) \ket{\alpha_0, \xi}$. Any such potential numerical approach must include the contribution of the squeezing, $\xi$, into the dipole spectrum, $d(\omega)$, such that these effects are noticeable in the HHG energy spectrum, $d\mathcal{E} / d\omega$. We will now develop two methods that are in principle capable of describing quantum light effects in HHG, and compare their effects on the dipole spectrum $d(\omega)$: an exact single-mode quantum electrodynamical model (Sec.~\ref{sec: qed}) and an approximate semiclassical multi-trajectory model (Sec.~\ref{sec: mtef}). Notably both models presuppose that photons live in a cavity, which enables the use of quantized photon modes and relies on methods developed for {\it ab initio} cavity electrodynamics~\cite{cohen-qed, motion-charged-particles-BSV, spontaenous-emission-multitrajectory, high-photon-entangled-coherent, principle-nonlinear-optics-mukamel, QEDFT-rugge, QEDFT-rugge-2, QEDFT-rugge-3}.

\subsection{Single-mode quantum electrodynamical model}
\label{sec: qed}

The Hamiltonian of our single quantized photon mode model consists of an electronic Hamiltonian $\hat{H}_A$, a single photon mode Hamiltonian $\hat{H}_F$, and the length-gauge dipole approximation interaction Hamiltonian $\hat{H}_{\mathrm{int}}$~\cite{motion-charged-particles-BSV, Entanglement-squeezing} (we use atomic units unless stated otherwise):

\begin{equation}
    \hat{H} = \hat{H}_A + \hat{H}_F + \hat{H}_{\mathrm{int}},
    \label{eq: hamiltonian}
\end{equation}

\noindent where the electronic Hamiltonian is that of a one-dimensional model potential $\hat{H}_A = \hat{\nabla}_x^2/2 - 1/\sqrt{\hat{x}^2 + b^2} + \lambda^2 \hat{x}^2 /2$, with $b$ the softening parameter and $\lambda$ the light-matter coupling (this coupling defines the cavity length $\lambda = \sqrt{8\pi/L_C}$ and physically describes the amplitude of vacuum fluctuations inside the cavity~\cite{vacuum-fluctuations-coupling}). The term $\lambda^2 \hat{x}^2/2$ describes the self-interaction of the electronic dipole through the photon-mode in the length gauge~\cite{cohen-qed, principle-nonlinear-optics-mukamel}. The free photon Hamiltonian $\hat{H}_F = \hat{\nabla}_y^2 /2 + \omega_L^2 \hat{y}^2 /2$ is that of a simple harmonic oscillator, where $\omega_L$ is the photon mode frequency. The operator $\hat{y}$ is related to the creation and annihilation operators of the photon mode via $\hat{y} = \left( \hat{a} + \hat{a}^\dagger \right)/ \sqrt{2 \omega_L}$. Note that the free photon Hamiltonian can also be rewritten in terms of the creation and annihilation operators as $\hat{H}_F = \omega_L (1/2 + \hat{a}^\dagger \hat{a})$. Finally, the length gauge interaction Hamiltonian is the dipole coupling to the electric field $\hat{H}_{\mathrm{int}} = - \hat{x} \hat{E}(t)$. The electric field operator of the photon is:
\begin{equation}
    \hat{E}(t) = \lambda \omega_L f(t) \hat{y} = \lambda f(t) \sqrt{ \frac{\omega_L}{2} } \left[ \hat{a} + \hat{a}^\dagger  \right],
    \label{eq: efield}
\end{equation}

\noindent such that the interaction Hamiltonian takes the form $\hat{H}_{\mathrm{int}} = - \lambda \omega_L f(t) \hat{x} \hat{y}$ if we write it in terms of the position operators $\hat{x}$ and $\hat{y}$. In this representation overall the electronic coordinate is represented by the coordinate $x$, and the photonic coordinates by $y$. In Eq.~\eqref{eq: efield}, $f(t)$ is the envelope of the light-matter interaction, which is off at the beginning of the simulation $f(t_0) = 0$ and is smoothly turned on during the simulation time to $f(t) = 1$, as is back off at the end of the simulation $f(t_0+T) = 0$. The specific shape of $f(t)$ (depicted in Fig.~\ref{fig: figure-paper}) used in our simulation is:

\begin{equation}
    \begin{split}
    f(t_0+t) = \left[ 1-\Theta(t - \tau_{\mathrm{1}})\right]\sin^2{\left( \pi t / 2 \tau_{\mathrm{1}} \right)} + \Theta(t - \tau_{\mathrm{1}}) -\\
    -~\Theta(t - \tau_{\mathrm{2}}) + \Theta(t-\tau_{\mathrm{2}})\sin^2{\left\lbrace \pi \left[ (t-\tau_{\mathrm{2}}) / \tau_{\mathrm{1}} +1\right] /2\right\rbrace},
    \label{eq: envelope}
\end{split}
\end{equation}

% Matan comment: huge \lambda -> try to simulate the atom in vacuum. Are these effects present for different values of lambda.

% Exact-factorization would consider \rho_A and \rho_F.

% \lambda* d << \sqrt{\omega_L}*\alpha can be numerically checked and its is consistent

% s < 1 corresponds to amplitude-squeezing. Consider writing s. Comment on non-quantum entanglement.

\noindent where $\tau_1 = 6\pi/\omega_L$ and $\tau_2 = 12\pi/\omega_L$, thus the total simulation time is $T = \tau_1 + \tau_2 = 18 \pi / \omega_L$, which translates into $9$ optical cycles with the full width at half maximum (FWHM) being 8 optical cycles. It corresponds to a trapezoidal shape with smooth transitions. As an initial state we choose the electronic ground-state, $\ket{g}$, and a squeezed-coherent state for the photon mode, $\ket{\alpha_0, \xi}$. The combined electron-photon initial state is then $\ket{ \Psi(t_0) } = \ket{g} \otimes \ket{\alpha_0, \xi}$. This model allows fully correlated light-matter wavefunctions as the system evolves over time. \\

\noindent We employ Octopus code~\cite{Octopus-1, Octopus-2} to solve the Schr\"odinger equation in this two-dimensional system $(x,y)$ by expressing photon modes as 1D harmonics oscillators in the photon phase-space [Eq.~\eqref{eq: efield}]. For a multimode photon system, it would suffice to increase the dimensionality of the problem by adding more coordinates $y^\prime, y^{\prime \prime}, ...$ to the Hamiltonian of Eq.~\eqref{eq: hamiltonian}. This has the disadvantage of exponential scaling, but enables us to benefit from already implemented and optimized  softwares that solve TDSE in N-dimensional systems, such as Octopus. At the very least, such an approach should be applicable for a single electronic coordinate, and up to five photonic ones, in accordance with efforts on exactly solving two-electron systems in 3D~\cite{2-electron-atom-6D}. Overall, this approach employs methods originally developed for quantum electrodynamics in cavities (e.g. for polaritonic chemistry~\cite{polaritonic-chemistry}), and repurposes it for quantum HHG by changing the boundary and initial conditions. This allows it to be implemented in typical quantum chemistry packages. \\

\noindent The numerical values of the parameters used in the simulation are the electron-photon coupling $\lambda = 0.015$ (corresponding to a photon cavity length of $L_C \sim 6~\mathrm{\mu m}$ and a cavity fundamental frequency of $\omega_C \sim 0.0038$ a.u.), the driving laser frequency $\omega_L = 0.057$ a.u. (corresponding to $\omega_L \sim 15 \omega_C$ and a wavelength of $\lambda_L = 800~\mathrm{nm}$), the coherent intensity of the squeezed-coherent states $\alpha_0 \sim 10.46$ (which corresponds to a maximum electric field intensity of $0.053$ a.u. or $10^{14}~\mathrm{W/cm}^2$), a squeezing parameter that is swept among a range of values $s = e^{2\xi} \in (0.05, 25)$, and the softening parameter for the electron model potential $b = 0.816$ a.u. (corresponding to a Neon ionization potential $I_p \sim 0.7925$ a.u.). The converged parameters of the simulation are those of a 2D time-dependent simulation over time: electron box size $L_x = 120$ a.u., photon box size $L_y = 100$ a.u., electron finite-difference step $dx = 0.7$ a.u., photon finite-difference step $dy = 0.1$ a.u., time-step $dt = 0.02$ a.u., and complex absorbing potential with absorbing length for both coordinates $L_{\mathrm{ab}} = 30$ a.u.~\cite{cap-umberto}. \\

\noindent Note that the numerical parameters are converged for this particular value of $\lambda$, as the light state is weighted by $\lambda$ when analyzing the scale of the photon coordinate $y$ [see Eq.~\eqref{eq: efield} and~\eqref{eq: gaussian}], which also affects convergence. Although this choice of $\lambda$ is somewhat arbitrary, we will show below that the main minima feature in the HHG spectra exists for a wide range of values of $\lambda$, under the condition that a stronger coupling leads to a stronger effect of the squeezing in the HHG yield (see App.~A). The particular value we used corresponds to medium light-matter coupling as attainable within optical cavities~\cite{spontaenous-emission-multitrajectory}, and are expected to be physical for reasonable experimental conditions (though the exact choice of $\lambda$ heavily depends on the cavity geometry that the simulation employs~\cite{spontaenous-emission-multitrajectory, motion-charged-particles-BSV, quantum-optical-description-hhg, vacuum-fluctuations-coupling}). Importantly, $\lambda$ is not a fully independent parameter in our simulation, as together with the choice of $\alpha_0$ it defines the expectation value of the laser peak power. Thus, whenever different values of $\lambda$ are explored (see App.~A), we also set $\alpha_0$ to keep the same peak field strengh. %Ofer comment

\subsection{Multi-trajectory Ehrenfest description}
\label{sec: mtef}

Multi-trajectory Ehrenfest dynamics (MTEF) is a model that approximates a quantum electrodynamical simulation with multiple semiclassical simulations that take into account the quantum uncertainty as classical statistical uncertainty~\cite{spontaenous-emission-multitrajectory}, also analogous to including phonon modes~\cite{phonon-multitrajectory}. In our case, the photon mode is treated semiclassically while keeping the electronic system fully quantum, in-line with recent theories~\cite{spontaenous-emission-multitrajectory, benchmarking, phonon-multitrajectory}. MTEF approximates the comined electron-photon system to be uncorrelated, such that the combined density operator is $\hat{\rho}(t) \approx \hat{\rho}_A(t) \otimes \hat{\rho}_F(t)$, where $\hat{\rho}_A(t)$ represents the density operator for the electron (atomic) system and $\hat{\rho}_F(t)$ represents the density operator for the photon system. We expect improvements to the uncorrelated evolution between light and matter that MTEF presupposes, building upon recent work regarding the possibility of an exact-factorization procedure between light and matter~\cite{exact-factorization}. Ths contrasts with the single quantum photon mode of Sec.~\ref{sec: qed}, which grants that light-matter correlations are fully integrated. \\% Ofer comment

\noindent We use the Wigner representation for the photon system~\cite{Scully_Zubairy_1997}, such that the photonic density operator becomes a function of the complex phase-space variable $\alpha$ (usually interpreted as the classical phase-space variable). The Wigner distribution of the photon mode is the analog of the density operator in the Wigner representation: $\hat{\rho}_F(t) \rightarrow \rho_{F, W}(\alpha; t)$, with $\rho_{F,W}(\alpha; t) \equiv \braket{\hat{\delta}(\hat{a} - \alpha)}(t) = \mathrm{Tr} \left[ \hat{\rho}_F(t) \hat{\delta} (\hat{a} - \alpha) \right]$. Expectation values are expressed as integrals over the phase-space variable $\alpha$: $\braket{\hat{\mathcal{O}}} \rightarrow \int d^2\alpha \mathcal{O}(\alpha) \rho_W(\alpha)$, where $\mathcal{O}(\alpha)$ is the operator $\hat{O}$ in the Wigner representation~\cite{Scully_Zubairy_1997}. In the multi-trajectory Ehrenfest algorithm, we approximate the Wigner distribution of the photon at the initial time $\rho_{F, W}(t_0)$ to be equivalent to a statistical sampling of $N_{\mathrm{traj}}$ trajectories $\alpha^j$, where $j$ is the index of the trajectory, enabling a classical treatment of the photonic dynamics: 

\begin{equation}
    \rho_{F,W}(\alpha; t_0) \approx \sum_j \frac{1}{N_{\mathrm{traj}}} \delta^{(2)} \left[ \alpha - \alpha^j(t_0) \right].
\end{equation}

\noindent The electron-photon system density operator $\hat{\rho}(\alpha; t)$ (whose photon part is expressed in the Wigner representation) takes the form:~\cite{spontaenous-emission-multitrajectory, benchmarking}:

\begin{equation}
    \hat{\rho}_W(\alpha; t) = \frac{1}{N_\mathrm{traj}^2} \left\lbrace \sum_j \hat{\rho}_A^j(t) \right\rbrace \otimes \left\lbrace \sum_j \delta^{(2)} \left[ \alpha - \alpha^j(t) \right] \right\rbrace,
    \label{eq: MTEF density}
\end{equation}

\noindent where $\rho^j_A (t) = \ket{\psi^j(t)}  \bra{\psi^j(t)}$ are the electronic wavefunctions that evolve dynamically with the classical photon mode $\alpha^j(t)$ for each respective trajectory $j$. The dynamical evolution of the coupled electron and photon system is given by the semiclassical set of equations [in contrast to the fully quantum Hamiltonian of Eq.~\eqref{eq: hamiltonian}]~\cite{spontaenous-emission-multitrajectory, benchmarking, cohen-qed, principle-nonlinear-optics-mukamel}:

\begin{equation}
\left\{
\begin{split}
    i \frac{\partial}{\partial t} \ket{\psi^j(t)} = \left[ \hat{H}_A - \hat{x} \sqrt{2 \omega_L} \lambda \Re \left\lbrace \alpha^j(t) \right\rbrace \right] \ket{\psi^j(t)}, \\
    \dot{\alpha}^j(t) + i\omega_L \alpha^j(t) = i\lambda \sqrt{\frac{\omega_L}{2}} \bra{\psi^j(t)} \hat{x} \ket{\psi^j(t)},
\end{split}
\label{eq: mxwll-TDSE}
\right.
\end{equation}

\noindent where $\hat{H}_A$ is the Hamiltonian for the electronic system, which is the same as in Eq.~\eqref{eq: hamiltonian}. The initial conditions for each photon trajectory $\alpha^j(t_0)$ are sampled from the Wigner distribution of the initial photon state $\rho_{F,W}(\alpha; t_0)$. We will assume that there is no feedback from the electronic system into the classical photon trajectories, that is, $\lambda\braket{\hat{x}}(t) \ll \sqrt{\omega_L} \alpha^j(t)$ in Eq.~\eqref{eq: mxwll-TDSE}. The evolution of the photon modes becomes that of free Maxwell equations: $\alpha^j(t) \approx \alpha^j(t_0) e^{-i \omega_L (t - t_0)}$. By substituting this into Eq.~\eqref{eq: mxwll-TDSE}, we finally reach the TDSE for the electornic system coupled to the classical photon mode: 

\begin{equation}
    i \frac{\partial}{\partial t} \ket{\psi^j(t)} = \left[ \hat{H}_A - \hat{x} E_j f(t) \cos{\left[ \omega_L (t - t_0) + \phi_j \right]} \right] \ket{\psi^j(t)},
    \label{eq: TDSE}
\end{equation}

\noindent where $E_j = \sqrt{2 \omega_L} \lambda | \alpha^j(t_0) |   $ and $\phi_j = -\arg \left[ \alpha^j(t_0) \right]$ are the field amplitude and phase, respectively, for the corresponding trajectory $j$, with $E_j(t) = E_j f(t) \cos{\left[ \omega_L (t - t_0) + \phi_j \right]}$ the electric field for this trajectory. The expectation values of the electron operators can be computed via the average response from the individual and independent trajectories through Eq.~\eqref{eq: MTEF density}:

\begin{equation}
\begin{split}
    \braket{\hat{\mathcal{O}}_A}(t) = \int d^2\alpha\mathrm{Tr} \left[ \hat{\mathcal{O}}_A \hat{\rho}(\alpha; t) \right] = \\
    = \frac{1}{N_{\mathrm{traj}}} \sum_j \bra{\psi^j(t)} \hat{\mathcal{O}}_A \ket{\psi^j(t)}.
\end{split}
\label{eq: expectation}
\end{equation}

\noindent For squeezed-coherent states, the Wigner distribution has an analytical expression which corresponds to a gaussian distribution $W_{\alpha_0 \xi}(\alpha)$ defined by the parameters of its coherence and squeezing, $\alpha_0$ and $\xi$ respectively~\cite{Scully_Zubairy_1997, BSV-husimi}:

\begin{equation}
    W_{\alpha_0 \xi}(\alpha) = \frac{1}{\pi} \exp{\left\lbrace e^{2\xi} \left( \Re \left[ \alpha \right]  - \alpha_0 \right)^2 + e^{-2\xi} \Im \left[ \alpha \right] ^2  \right\rbrace}.
    \label{eq: gaussian}
\end{equation}

%\noindent The gaussian of Eq.~\eqref{eq: gaussian} is centered at $\alpha = \alpha_0$ and with standard deviation in the axis of  axis $e^{-2\xi}$ for the real part and $e^{-2\xi}$ for the imaginary part.  
\noindent The squeezing parameters $\xi$ and $s = e^{2\xi}$  will be used interchangeably throughout the manuscript. Three cases of $\xi$ are relevant for this work: $\xi < 0$ or $s < 1$ correspond to amplitude-squeezing, $\xi = 0$ or $s = 1$ correspond to no-squeezing (coherent state), and $\xi > 0$ or $s > 1$ correspond to phase-squeezing. Since we will use squeezed-coherent states as the intial photon states in our simulation, $\hat{\rho}_F(t_0) = \ket{\alpha_0, \xi}\bra{\alpha_0, \xi}$, the initial Wigner distribution of our photon modes is defined by Eq.~\eqref{eq: gaussian}: $\rho_{F,W}(\alpha; t_0) = W_{\alpha_0 \xi}(\alpha)$, which will be used for the sampling of the initial values of the photon phase-space variable $\alpha^j(t_0)$. \\ 

\noindent The single quantized photon mode runs in approximately $20$ minutes on one CPU, while the multi-trajectory simulations requires $30$ seconds per trajectory on one CPU, and the MTEF simulation reaches convergence at around $10000$ trajectories. The relative efficiency of MTEF compared to full QED simulation relies on the easy possibility to avoid the expontential scaling of the bosonic basis sets (already for two photon modes, we expect MTEF to be substantially faster and less computationally heavy than solving the quantum dynamics exactly).

\begin{figure}[ht]
    \centering
    \includegraphics[width=\linewidth]{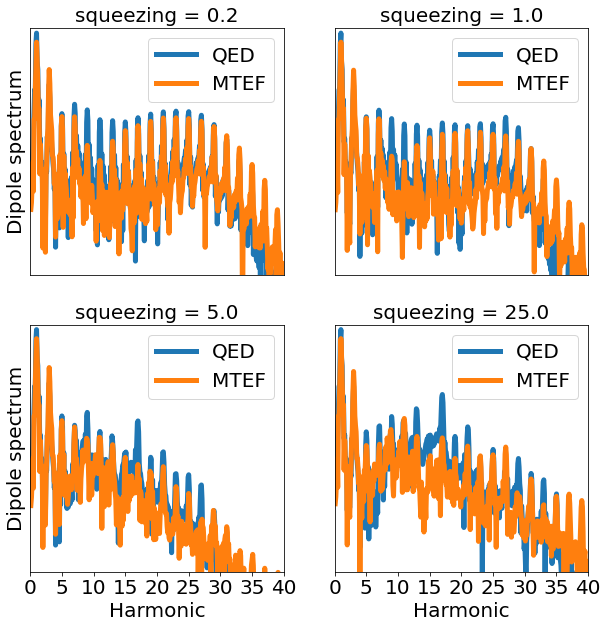}
    \caption{HHG emission spectra for different values of the squeezing parameter $s = e^{2\xi} = 0.2$, $1.0$, $5.0$, and $25.0$; and comparing both methods QED (blue) and MTEF (orange). As we enter into the phase-squeezing regime $s>1$ a loss of the typical plateau ($11<n<31$) is observed, predicted by both methods (QED and MTEF).}
    \label{fig: spectrum-squeezed}
\end{figure}

%%%%%%%%%%%%%%%%%%%%%%%%%%%%%%%%%%%%%%%%%%%%%%%%%%%%%%%%%%%%%%%%%%
%                       Computational results                    %
%%%%%%%%%%%%%%%%%%%%%%%%%%%%%%%%%%%%%%%%%%%%%%%%%%%%%%%%%%%%%%%%%%
\section{Computational results and discussion}
\label{sec: results}

%\subsection{Comparison between the quantum simulation and MTEF}
% Ofer comment
The observable to be compared between the two simulations is the dipole moment of the electron $d(t)$ for different squeezings $\xi$, and the resulting HHG spectra, computed as an expectation value of the electron-photon time-dependent wavefunction for the QED model: $d_{\xi}(t) = \bra{ \Psi_{\xi}(t) } \hat{x} \ket{\Psi_{\xi}(t)}$, where $\ket{\Psi_{\xi}(t_0)} = \ket{g} \otimes \ket{\alpha_0, \xi} $; and using a sum of expectation values for MTEF: $d_{\xi}(t) = \sum_j \bra{\psi^j(t)} \hat{x} \ket{\psi^j(t)}/N_{\mathrm{traj}}$ [Eq~\eqref{eq: expectation}], where the phase-space variable for each trajectory, $\alpha^j(t_0)$ [see Eq.~\eqref{eq: TDSE}], has been sampled from a Wigner distribution of its corresponding squeezing $W_{\alpha, \xi}(\alpha)$ [Eq.~\eqref{eq: gaussian}] for MTEF. The dipole variable $d_\xi(t)$ is then Fourier transformed into the harmonic spectrum $d_{\xi}(\omega)$ also masked by the envelope of the incident field~\cite{Entanglement-squeezing}:

\begin{equation}
    d_{\xi}(\omega) = \int_{t_0}^{t_0+T}  d_{\xi}(t) f(t) e^{-i \omega t} dt.
\end{equation}

\noindent Figure~\ref{fig: spectrum-squeezed} presents the calculated HHG emission from both methods in typical conditions, and for different values of the squeezing parameter $s = e^{2\xi} = 0.2$, $1.0$, $5.0$, $25.0$. The data in Fig.~\ref{fig: spectrum-squeezed} predicts that phase-squeezing removes the plateau ($11 < n< 31$) and, instead, manifests a consistent drop in its HHG yield (a phenomenon observed already in previous semiclassical works~\cite{matan-coherent-squeezed}). Figure~\ref{fig: spectrum-squeezed} reveals some discrepancy between MTEF and QED, which suggests true quantum effects beyond semiclassical interpretations play a role in the dynamics and HHG emission mechanism. \\

\begin{figure}[ht]
    \centering
    \includegraphics[width=\linewidth]{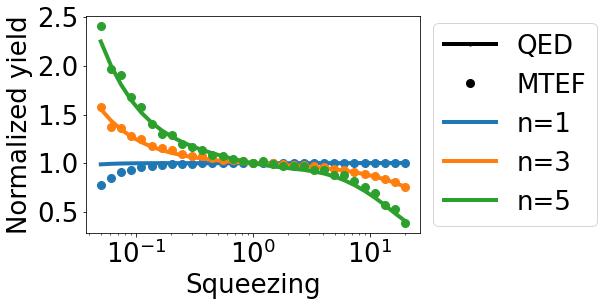}
    \caption{Normalized yield for the first [$y_1(\xi)$], third [$y_3(\xi)$], and fifth harmonics [$y_5(\xi)$] for the quantum electrodynamical simulation (QED, solid lines) and the multi-trajectory Ehrenfest dynamics (MTEF, circles)  vs. the squeezing parameter $s = e^{2\xi}$.}
    \label{fig: low-harmonics}
\end{figure}

\begin{figure*}[ht]
    \centering
    \includegraphics[width=\linewidth]{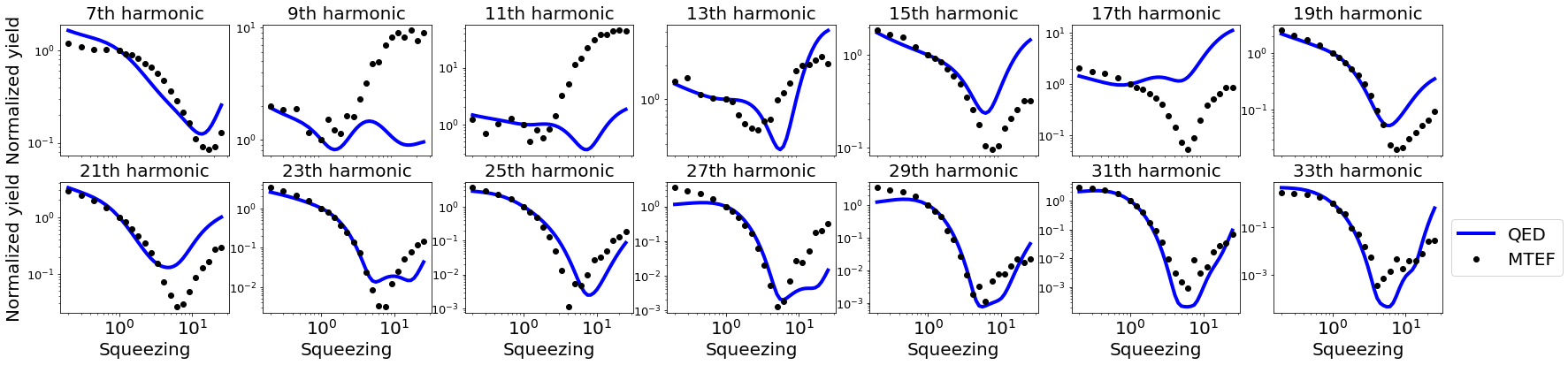}
    \caption{Normalized HHG yield for the quantum electrodynamical simulation (QED, solid blue line) and the multi-trajectory Ehrenfest dynamics (MTEF, black circles) from the seventh [$y_7(\xi)$] to the $33$th harmonic [$y_{33}(\xi)$]  vs. the squeezing parameter defined as $s = e^{2\xi}$. In all of the harmonics we observe a minimum at a phase-squeezing value $s > 1$ ($\xi > 0$) which is captured by both methods.}
    \label{fig: yield-harmonics}
\end{figure*}

\noindent The spatial symmetry of the electronic model potential together with that of the incident light forbids even harmonics of $\omega_L$ from being emitted~\cite{selection-rules-1, qo-nature-hhg, Theory-hhg}. Consequently, for analyzing the HHG yield we integrate the dipole spectrum from even harmonic to even harmonic to get the harmonic-order-resolved yield $Y_{2n+1}(\xi) \equiv \int_{2n\omega_L}^{(2n+2)\omega_L} \omega^4 | d_{\xi}(\omega) |^2 d\omega$ (Fig.~\ref{fig: figure-paper}). We compare the normalized yield $y_n(\xi)$ for different squeezings, that is, the yield of the dipole emission for different harmonics $Y_n(\xi)$ normalized to the dipole emission of the coherent state $Y_n(\xi=0)$: $y_n(\xi) = Y_n(\xi)/Y_n(0)$. The computational results comparing the quantum electrodynamical simulation and the multi-trajectory Ehrenfest dynamics are shown in Fig.~\ref{fig: low-harmonics} and~\ref{fig: yield-harmonics} for each of the harmonics from the first to the $33$th (the cutoff is at around $n_{\mathrm{cutoff}} \sim 27$), where we see the normalized yield $y_n(\xi)$  vs. squeezing $s = e^{2\xi}$. $y_n(\xi)$ is particularly hard to test for MTEF (as opposed to simpler observables such as forbidden harmonics, cut-off scalings, etc), but it is an interesting variable to analyze effects of squeezing in HHG. \\

\begin{figure}[ht]
    \centering
    \includegraphics[width=\linewidth]{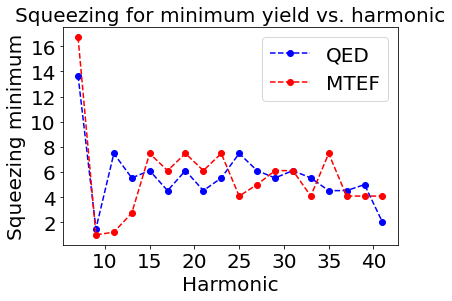}
    \caption{Squeezing $s = e^{2\xi}$ at which the yield is minimum vs. harmonic order for both simulations QED (blue) and MTEF (red).}
    \label{fig: minimum}
\end{figure}

\noindent We observe from Fig.~\ref{fig: low-harmonics} that the changes in the lower perturbative harmonics due to squeezing (solid line) are fully captured by the statistical distribution of the initial squeezed-coherent state (round dots), which means that the effects of the squeezing in these harmonics can be explained using the initial Wigner distribution of Eq.~\eqref{eq: gaussian} through MTEF. This suggests that lower harmonics can be described by hidden-variables methods. Figure~\ref{fig: low-harmonics} HHG predicts, in addition, that there is an increase in the yield for amplitude-squeezing and a decrease in yield for phase-squeezing.  \\
% Matan comment: We have both phase- and amplitude-squeezing

\noindent Figure~\ref{fig: yield-harmonics} presents the main physical results of this manuscript analyzing the HHG yield for higher order harmonics vs. squeezing: (i) We numerically observe a clear spectral minima behaviour vs. squeezing that is universal for all harmonic orders (see Fig.~\ref{fig: minimum}). This feature and the exact minima position is expected to be highly sensititve to the system parameters (e.g. laser regime and electronic structure), which make it potentially useful for developing novel ultrafast quantum spectroscopies. (ii) The MTEF simulation only partially reconstructs this minima structure (e.g. in harmonics $19 - 33$, but failing in harmonics $11$ and $17$, and often missing the exact squeezing value for which the minima is obtained). Remarkably, this disagreement between MTEF and QED means that the approximations in MTEF are likely too strong for HHG driven by squeezed light, and hints that true entanglement (not included in hidden-variable theories) plays a role in HHG emission~\cite{Cauchy-Schwarz}. Let us further emphasize that the minima structure is also attained for almost all harmonics in a wider range of coupling strength (see App.~A), but the position of the minima differs between chosen parameters. \\ %Ofer comment
%erase some relevant phenomena, such as the absence of light-matter entanglement that was presupposed in MTEF and which no hidden-variables method can capture due to Bell's theorem~\cite{Cauchy-Schwarz}, an effect which would be captured by the QED model (maybe future improvements of MTEF are able to capture this effect~\cite{exact-factorization}).

% Matan comment: plot minimum vs order
% Matan comment: I have to investigate over Schmidt decomposition and harmonics
\noindent In further analysis we find that the squeezing value at which the HHG minima is found is numerically close to the minima of the instantaneous correlation functions of the photon initial states $g^{(n)}(\xi) = \bra{\phi_\xi}\hat{a}^{\dagger n} \hat{a}^{n} \ket{\phi_\xi}/\bra{\phi_\xi} \hat{a}^\dagger \hat{a} \ket{\phi_\xi}^{n}$ for $\ket{\phi_\xi} = \ket{\alpha_0, \xi}$, which might be explained by the multiphoton proceses that are involved in the dipolar emission of these harmonics~\cite{Correlation-function-experiment}. However, we found that this value does not match the scaling of the minima position with the laser parameters such as intensity and wavelength (see App.~B), meaning it likely does not capture the physical mechanism causing it. \\ 

\noindent Two harmonics are especially important for the analysis: $n =9$ and $n=11$. These harmonics are resonant with the electronic system (close to a transition in the electronic system) and this might explain the much more pronounced difference between MTEF and exact quantum dynamics for these energies as this resonance could be the source of high light-matter correlation, not covered in MTEF model~\cite{squeezing-from-resonance}. Thus, our numerical results that show a particularly strong disagreement between MTEF and full quantum simulations for resonance harmonics suggest potential strong quantum optical effects in molecular or atomic resonances~\cite{giant-Xe-resonance, HHG-shape-resonance, HHG-resonance, mol-anisotropy-shape-response, discriminating-phase-matching}. Our results reveal the capacity of MTEF to predict many of the qualitative effects that squeezed light causes on the electronic dipole, but they simultaneously also expose the limitation of the current description of MTEF to cover all the quantitative effects observed in fully QED simulations.

\section{Conclusion}
\label{sec: conclusion}

% rephrase the conclusion, think whether to include the entanglement graph

%We tested quantum HHG simulations comparing between two different methods for ultrafast electronic dynamics under a strong and squeezed driving field: a quantum electrodynamical model using a single quantized photon mode for the driving field, and an approximate semiclassical multi-trajectory Ehrenfest simulation. We tested MTEF semi-classical approximation for quantum HHG and whether is capable of capturing the squeezing-dependence of the HHG yield and provide a milestone on how HHG with quantum light could be tested in future sytems. Figure~\ref{fig: spectrum-squeezed} reveals that phase-squeezing more specific the characteristic plateau-cutoff structure of the HHG in the dipole and, instead, shows a consistent drop in the harmonic intensity, a phenomenon that both methods (QED and MTEF) predict. MTEF is able to explain many of the changes that occur in the HHG yield due to the squeezing, e.g. the existence of a characteristic minima structure, and especially the behaviour of perturbative harmonics. However, we additionally find that not all of the electronic dynamics of the exact model can be explained by our MTEF simulations, which some true quantum effects are required for explaining the full the electron-photon interaction. This result is expected to be improved in the future by including interactions between the trajectories in MTEF, which would potentially capture some light-matter correlation effects that are missing in our current description. \\

We explored quantum HHG simulations comparing two different methods for ultrafast electronic dynamics under a strong and squeezed driving field: an exact quantum electrodynamical model using a single quantized photon mode for the driving field, and an approximate semiclassical multi-trajectory Ehrenfest simulation. We tested MTEF semi-classical approximation for quantum HHG, concluding that it partially captures the squeezing-dependence of the HHG yield. MTEF is able to explain many of the changes that occur in the HHG yield due to the squeezing, e.g. the existence of a characteristic minima structure, and especially the behaviour of perturbative harmonics. This result provides a milestone on how HHG with quantum light could be tested in future systems, and paves the way to a universal framework of QED-HHG. Moreover, our result reveals that phase-squeezing qualitatively affects the HHG spectrum by removing the characteristic plateau structure, manifesting instead an irregular pattern of decreasing HHG yield, a phenomenon that both methods (QED and MTEF) predict. However, we additionally find that not all of the HHG spectral features can be explained by our MTEF simulations, which hints that true quantum effects are required for explaining the full the electron-photon interaction. In particular, the exact position of HHG minima vs. squeezing is sensitive to the level of theory, suggesting it could provide an emerging observable in novel ultrafast quantum spectroscopies, as well as to benchmark new theories and approximations (especially near resonances that could serve as novel platforms for entanglement~\cite{giant-Xe-resonance, HHG-shape-resonance, HHG-resonance, mol-anisotropy-shape-response, discriminating-phase-matching}). Looking forward, our work should motivate further theoretical developments and proposes an experimental set-up and test to benchmark theory and uncover quantum effects in HHG. 
%To sum up, this work validates, to a certain extent, using semiclassical techniques in the field of nonlinear quantum optics, which have already been successful in other contexts like quantum chemistry and cavity QED. This is a crucial conclusion for this growing field, since semiclassical approaches have been thus far employed ad-hoc, and they are the main technique which might be able to include many active quantized photon modes in future simulations (given the exponential scaling of the system size). At the same time, our work cautions the untested use of such schemes, as certain features are not always well described, such as the specific position of HHG minima, possibly suggesting substantial light-matter entanglement. Our work proposes a roadmap for {\it ab initio} predictions for HHG in realistic setups by employing methods compatible with existing quantum chemistry software.

\section*{Acknowledgements}

This work was supported by the European Research Council (ERC-2015-AdG694097), the Cluster of Excellence ‘Advanced Imaging of Matter’ (AIM), Grupos Consolidados (IT1453-22) and Deutsche Forschungsgemeinschaft (DFG) - SFB-925 - project 170620586. The Flatiron Institute is a division of the Simons Foundation. We acknowledge support from the Max Planck-New York City Center for Non-Equilibrium Quantum Phenomena. S.d.l.P. acknowledges support from International Max Planck Research School. The authors also would like to acknowledge the computational support provided by Max Planck Computing and Data Facility. We would like to thank Michael Ruggenthaler and Mark Kamper Svendsen for some interesting discussions. 

%%%%%%%%%%%%%%%%%%%%%%%%%%%%%%%%%%%%%%%%%%%%%%%%%%%%%%%%%%%%%%%%%%
%                        Bibliography                            %
%%%%%%%%%%%%%%%%%%%%%%%%%%%%%%%%%%%%%%%%%%%%%%%%%%%%%%%%%%%%%%%%%%
\bibliography{references} % Produces the bibliography via BibTeX.
%\printbibliography

\appendix

\section{Light-matter coupling analysis}
\label{sec: lambda}

The light-matter coupling $\lambda$ of Eq.~\eqref{eq: efield} and~\eqref{eq: mxwll-TDSE} is a parameter that we presuppose in our model, as we are using a single photon mode in a cavity. We use the value $\lambda = 0.015$ in accordance with ref.~\cite{spontaenous-emission-multitrajectory}. This section is meant to provide a brief discussion on the consequences of changing the value of $\lambda$. For our 1D cavity modes, the value of the light-matter coupling is related to the size of the cavity $L_c$ through: $\lambda = \sqrt{8 \pi / L_C}$~\cite{spontaenous-emission-multitrajectory, benchmarking} (analogous to the quantization volume in 3D models~\cite{cohen-qed, motion-charged-particles-BSV, quantum-optical-description-hhg, principle-nonlinear-optics-mukamel}) \\

\noindent Figure~6 presents examplary HHG spectra for various values of $\lambda$ (changing by $\pm 33 \%$ for the value in the main text). The HHG spectra is largely unaffected by these changes, supporting the generality of our conclusions, and indicating our predictions would hold in a wide range of experimental conditions. At very high squeezing values (Fig.~6 for $s = 25.0$), stronger changes begin to emerge in the harmonic spectra, especially beyong the cutoff. This agrees with the semiclassical interpretation of the photon mode Wigner distribution [Eq.~\eqref{eq: gaussian}], since a higher $\lambda$ means that the effects of the squeezing on the matter system are amplified, and thus for a given value of phase-squeezing we obtain higher electric field fluctuations (they scale as $\Delta E_{\mathrm{max}} \sim \lambda e^{\xi}$) in relation to the laser amplitude that remains constant at $E_0 = 0.053~\mathrm{a.u.}$. We observe, nontheless, that the HHG minimum vs. squeezing is still present for these values of the light-matter coupling for almost all harmonic orders. In real measurements, the value of $\lambda$ should be determined experimentally.

%Figure~6 presents the HHG spectrum using three values for the light-matter coupling we named strong ($\lambda = 0.020$), normal ($\lambda = 0.015$), and weak ($\lambda = 0.010$) couplings. We observe that the choice of the coupling value does change the phase-squeezed HHG spectrum, and that this affects, among other things, the position of the yield minimum (as the coupling $\lambda$ grows larger, the yield minimum requires lower phase-squeezing) and the relative decrease of the HHG yield at this minimum (as can be observed in Fig.~6 when the squeezing is $s = 5.0$). 

% Matan comment: I cannot do that unless I have p in the Hamiltonian

\begin{figure*}[ht]
    \centering
    \includegraphics[width=\linewidth]{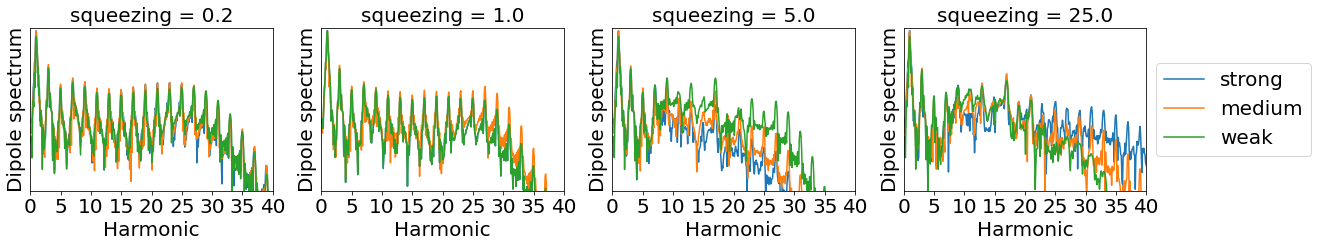}
    \caption{HHG yield using three values of the light-matter coupling $\lambda$: strong coupling $\lambda = 0.020$, medium coupling $\lambda = 0.015$ and weak coupling $\lambda = 0.010$. Four values of the squeezing are given $s = e^{2\xi} = 0.2$, $1.0$, $5.0$, $25.0$.}
    \label{fig: coupling-figure}
\end{figure*}

\section{Testing the correlation function hypothesis}
\label{sec: correlation}

The minima structure observed for most of the harmonics in Fig.~\ref{fig: yield-harmonics} motivated an analysis of what exactly was causing this phenomenon. It turns out that this minima structure is observed also in the nth-order instantaneous correlation function of the initial squeezed-coherent state $g^{(n)}(\xi)$ vs. the squeezing of this state for a fixed $\alpha_0$~\cite{Scully_Zubairy_1997}:

\begin{equation}
    g^{(n)}(s) = \frac{\bra{\alpha_0, \xi} \hat{a}^{\dagger n} \hat{a}^n \ket{\alpha_0, \xi}}{ \bra{\alpha_0, \xi} \hat{a}^{\dagger} \hat{a} \ket{\alpha_0, \xi}^n},
    \label{eq: correlation}
\end{equation}

\noindent where $s = e^{2\xi}$. If we consider, for the sake of testing this hypothesis, running the single quantum photon mode simulation for various laser intensities (correspoding to varing $\alpha_0$) we can obtain the resulting minima in the HHG yield as we did in Fig.~\ref{fig: yield-harmonics} (which would be the case $\alpha_0 = 0.053$). We now compare the phase-squeezing value $s_{\mathrm{min}}$ at which this is observed for various $\alpha_0$ with the minimum of the correlation function in Eq.~(B1) for the corresponding initial state. The results of this comparison are shown in Fig.~7, where we observe that the trend of the HHG yield is not completely described by the correlation function, even though the numerical values and the increasing tendency vs. electric field intensity do have some qualitative agreement. This suggests that this hypothesis could still be ruled out in future research as a true mechanism to explain the results in HHG yield, even though it does reveal a numerical correlation for the parameters of our simulation, which has already been verified for lower harmonics in~\cite{Correlation-function-experiment}.

\begin{figure}[ht]
    \centering
    \includegraphics[width=\linewidth]{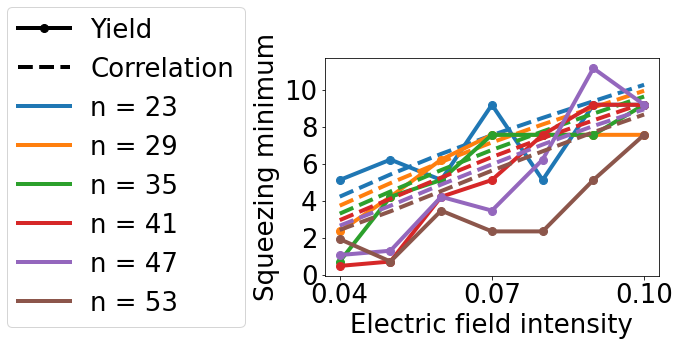}
    \caption{Comparison of the phase-squeezing minima $s_{\mathrm{min}}$ vs. the power of the laser $E_0 = \sqrt{2 \omega_L} \lambda \alpha_0$, for the single mode QED numerical simulation (dots) and the correlation function $g^{(n)}(s)$ defined in Eq.~(B1) [dashed line]. The number $n$ represents the yield harmonics we are considering and the order of the correlation function, respectively.}
    \label{fig: correlation}
\end{figure}

\end{document}